\documentstyle[aps,prb,multicol,psfig]{revtex}
\renewcommand{\narrowtext}{\begin{multicols}{2} \global\columnwidth20.5pc}
\renewcommand{\widetext}{\end{multicols} \global\columnwidth42.5pc}
\newcommand{\beq}{\begin{equation}}
\newcommand{\eeq}{\end{equation}}
\newcommand{\bea}{\begin{eqnarray}}
\newcommand{\eea}{\end{eqnarray}}
\newcommand{\sone}{{\bf S1}}
\newcommand{\stwo}{{\bf S2}}

\newcommand{\AlGaAs}{\mbox{Al$_{x}$Ga$_{1-x}$As}}

\begin{document}

\draft

  \title{Optimal design and quantum limit for second harmonic
  generation in semiconductor heterostructures}

  \author{Guido Goldoni\protect\cite{byline1}} 

  \address{Istituto Nazionale per la Fisica della Materia and \\
  Dipartimento di Fisica, Universit\`a di Modena, Via Campi 213/A,
  I-41100 Modena, Italy}

\date{\today}
\maketitle

\begin{abstract}

  The optimal design for infrared second harmonic generation (SHG) is
  determined for a GaAs-based quantum device using a recently
  developed genetic approach. Both compositional parameters and
  electric field are simultaneously optimized, and the quantum limit
  for SHG, set by the trade-off between large dipole moments
  (favouring electron delocalization) and large overlaps (favouring
  electron localization), is determined. Optimal devices are generally
  obtained with an asymmetric double quantum well shape with narrow
  barriers and a graded region sideways to the largest well. An
  electric field is not found to lead to improved SHG if compositional
  parameters are optimized.

\end{abstract}


\narrowtext

\section*{Introduction}

  In semiconductor quantum structures intersubband transitions lead to
  very large second-order optical susceptibilities \cite{Rosencher89};
  in particular, asymmetric quantum confinement may induce large
  second harmonic generation (SHG) in material systems, like Si/SiGe,
  where SHG is prohibited by symmetry in the centro-symmetric bulk
  host materials \cite{Seto94}; in non-centrosymmetric systems, like
  GaAs/AlGaAs, quantum confinement-induced SHG may exceed by orders of
  magnitude the bulk contribution
  \cite{Boucaud90,Rosencher91,Bewley93,Rosencher93}. Intersubband SHG
  has been intensely investigated in these and other classes of
  materials, like AlInAs/GaInAs \cite{Sirtori91,Sirtori92} or
  GaN/AlGaN \cite{Liu00}; band-to-band non-linear susceptibility has
  also been studied \cite{Khurgin,Kuwatsuka94,Park99}

  The intersubband SHG coefficient is given by\cite{Rosencher91}
\bea\label{chi}
\chi^{(2)}_{2\omega} & = & \frac{q^3}{\epsilon_0} \sum_{ij}
\frac{1}{2\hbar\omega+E_{ij}-i\Gamma_{ji}}
\sum_k \mu_{ij}\mu_{jk}\mu_{ki} \times \nonumber\\
& & \left[\frac{\rho_i-\rho_k}{\hbar\omega+E_{ik}-i\Gamma_{ki}}-
\frac{\rho_k-\rho_j}{\hbar\omega+E_{kj}-i\Gamma_{jk}}\right]
\eea
  where $E_{lm} = E_l-E_m$ is the transition energy between subbands
  $l$ and $m$, $\mu_{lm} = \langle l | z | m \rangle$ is the dipole
  matrix element of the transition, $\hbar\omega$ is the photon
  energy, $\rho_l$ is the surface charge density of subband $l$, and
  $\Gamma_{lm}$ accounts for inhomogenous broadening of the
  $l\rightarrow m$ transition. Most studies of SHG focus on the
  double-resonant regime, where $\hbar\omega$ is in resonance with two
  intersubband gaps; in this regime two denominators in (\ref{chi})
  are simultaneously resonant, and the device can be scketched as a
  three level system with the frequency of the incoming
  electromagnetic field matching approximately the $1\rightarrow 2$
  and $2\rightarrow 3$ transitions. Hence, at the resonant frequency
  $\omega_r$, $\chi^{(2)}_{2\omega_r} = q^3 \rho_1/(\epsilon_0
  \Gamma^2) \xi(\omega_r)$, where
\beq\label{xi}
\xi(\omega) = \frac{\mu_{12}\mu_{23}\mu_{31} \Gamma^2}
{(2\hbar\omega+E_{13}-i\Gamma)(\hbar\omega+E_{12}-i\Gamma)}
\eeq

  Quantum confinement-induced SHG (i.e., in addition to the possible
  bulk contribution) can only be observed if the confinement potential
  does not have the reflection symmetry, so that none of the dipole
  matrix elements in (\ref{xi}) is zero; this can be achieved either
  by an asymmetric composition
  \cite{Seto94,Boucaud90,Rosencher91,Bewley93,Shaw93} or by
  application of an electric field
  \cite{Liu00,Fejer89,Tsang88,Tsang92} $F$ or both
  \cite{Sirtori91}. However, the use of an electric field, in addition
  to a built-in asymmetry, does not necessarily improve SHG provided
  that resonance is achieved by proper design of the heterostructure,
  so that in general the usefulness of an applied electric field from
  this point of view has still to be assessed.

  Double-resonant SHG can be maximized by appropriately engineering
  the structure, in order to make the product of the three dipole
  matrix elements appearing in (\ref{xi}) as large as possible; to
  this end, the best compromise between large intersubband dipole
  moments (which favours electron delocalization) and large overlaps
  (which favours electron localization) must be found. The optimal
  design of the confinement potential, however, is a very hard
  optimization problem, since, in general, many compositional and
  geometrical parameters, and possibly the external field, can be
  varied, while resonance with a given radiation should be preserved
  in the process; furthermore, the optimal structure depends on the
  radiation frequency for a given class of materials. This
  optimization task has been undertaken so far only for specific
  confinement profiles; apart from the required asymmetry, the
  selected shapes were simple enough that optimization could be
  performed with respect to a single parameter\cite{Rosencher91} or
  the field \cite{Liu00,Tsang88,Tsang92}.  Analytical optimization
  methods for SHG have also been proposed \cite{Tomic97,Tomic98},
  providing a qualitative estimate of the optimal SHG which can be
  obtained in a heterostructure; however, such methods could only deal
  with idealized continuos potential profiles, and could not include
  physical limitations in the alloy concentrations; for example, the
  optimized potentials proposed in Refs.\
  \protect\onlinecite{Tomic97,Tomic98} cannot be implemented as
  strictly type-I structures.

  Recently, we have introduced a new numerical strategy which allows
  to optimize the performance of a given device with respect to all
  geometrical parameters and external fields {\em simultaneously}
  \cite{Goldoni00}. In short, our approach uses an artificial
  intelligence tecnique, namely, evolutionary programming\cite{EP}, to
  solve the inverse problem of designing a material with preset
  electronic properties, including complex non-linear constraints,
  such as multi-subbands transition energies; by exploring the
  multi-dimensional parameter space, the algorithm uses a genetic
  paradigm to find the regions providing the best performances, and
  focuses the search in those regions. Using a population-based
  strategy, and starting with a completely random initial set of
  potential profiles, the algorithm efficiently relaxes toward one or
  more optimal devices without limiting the search to a given class of
  structures.
  Details and demonstration of the algorithm can be found in Ref.\
  \onlinecite{Goldoni00}. The efficiency of our algorithm allows for a
  quantitative modelling of the structure, including, e.g., finite
  band-offsets, space-dependent effective masses, band
  non-parabolicity, and physical limitations on alloy concentrations.

  In this paper we apply our optimization strategy to determine the
  optimal composition for SHG from intersubband transitions in the
  conduction band of the GaAs/AlGaAs class of
  materials. We report the composition and
  potential profile for an optimal device with zero electric field. We
  also study optimal devices with a finite $F$, but we find that, in
  general, devices with a non-zero electric field have an estimated
  SHG which is weaker than devices with $F=0$.

  \section*{Details of the calculation}

  We consider an heterostructure composed of a large number of
  \AlGaAs\ layers, each consisting of $n$ monolayers (ML); in order to
  limit ourselves to type-I heterostructures, the Al molar fraction,
  $x$, can take continuous values in the range [0,0.4]; the
  GaAs/Al$_{0.4}$Ga$_{0.6}$As band offset is large enough to
  accomodate the three bound states which are required to give rise to
  double-resonant SHG. The structural parameters $x$ and $n$ of each
  layer and, in a sub-set of runs, also the eletric field $F$
  \cite{modified-qdoes}, were subjected to the evolutionary dynamics
  in order to maximize $\mu=|\mu_{12}\mu_{23}\mu_{31}|$ at the
  frequency of a CO$_2$ laser ($\hbar\omega = 0.116\,\mbox{eV}$).  As
  mentioned above, in the double-resonant regime the intersubband gaps
  match the radiation field; the possibility to implement such
  non-linear contraints is a distinct advantage of our numerical
  strategy. Since inhomogenous broadening lowers the peaks in the
  response function, our algorithm allows the structures to satisfy
  the resonance constraint only approximately; in the language of
  Refs.\ \onlinecite{Goldoni00,EP}, this is obtained using a ``penalty
  function'' with a Lorentzian shape and a width $\Gamma$ independent
  of the transition.

  Calculations were performed within a one-band envelope function
  description of the conduction band single-particle states in the
  low-density limit. The single-particle Hamiltonian, with a step-like
  potential determined by the sequence of Al concentrations \cite{BO}
  appearing at each step of the simulation, was represented in a
  plane-wave basis set normalized to a periodic box; capping layers of
  typically 40 MLs of Al$_{0.4}$Ga$_{0.6}$As terminate each strucure
  to ensure that the wavefunctions vanish at the boundaries. The
  space-dependent effective mass was linearly interpolated between the
  GaAs and AlAs masses for arbitrary Al concentrations
  \cite{EM}. Non-parabolic energy dispersion is known to have a
  sizable effect on subband energies and wavefunction localization for
  large confinement energies; this is of particular concern in the
  design of a device for SHG, where the three bound states involved
  span a large energy range above the conduction band edge. By
  explicit calculations, of the type described below, we have found
  that the optimal SHG intensiy can be overestimated by $\sim 30\,\%$
  if non-parabolicity is neglected; in our calculations
  non-parabolicity effects were taken into account using an
  energy-dependent effective mass\cite{NP}.

  \section*{Numerical results}

  The potential profiles and subband charge densities of an optimal
  structure, labelled \sone, obtained with $F=0$, and a sub-optimal
  structure, labelled \stwo, with $F\ne0$, are shown in Fig.\ 1. The
  corresponding compositional parameters are reported in Tab.\ 1.  As
  can be seen from the figure, \sone\ has a remarkably simple shape,
  with sharp variations of the Al concentration rather than a graded
  composition (indeed, sub-optimal structures with a more graded
  composition were obtained during the simulations; see also
  \stwo). Structure \sone\ is remarkably similar to the step-graded
  device considered in Ref.\ \onlinecite{Rosencher91}, but {\em a)
  with a sharp (high and narrow) barrier between the two steps}, and
  {\em b) with a graded region sideways to the deeper well}, which
  allows the charge density of the third subband to slightly
  delocalize out of the well region; from a large set of runs, we
  conclude that {\em both these features are characteristics of the
  best performing structures}. The calculated optimal value $\mu=
  3.27\,\mbox{nm$^3$}$ of \sone\ exeeds by $37\,\%$ the value $ \mu =
  2.39\,\mbox{nm$^3$}$ estimated in Ref.\ \onlinecite{Rosencher91}.

  The structure \stwo\ is a
  typical example of a sub-optimal device with an optimized electric
  field substantially different from zero, which appeared in our
  simulations. The structure has a graded double quantum well shape;
  again, a graded region is present sideways to the largest well. As
  shown in Tab.\ 1, the estimated value $\mu = 3.126\,\mbox{nm$^3$}$
  at the optimal field $F = 43\,\mbox{kV/cm}$, which is $\sim 4\,\%$
  smaller than for \sone. 

  Indeed, calculations show that, in general, an electric field
  does not improve SHG if compositional parameters are optimized. In
  Fig.\ 2 we show the value of $\mu$ for 200 structures which are
  optimized with respect to structural parameters and the electric
  field (open dots); the devices \sone\ and \stwo\ are also indicated
  (solid triangles). As it is apparent, the best performing structures
  tend to have a small or negligible field; the possibility to achieve
  large SHG is substantially decreased for fields larger than
  approximately 60 kV/cm.

  Although the best performing device, \sone, is obtained for a null
  electric field, an optimally designed biased heterostructure may be
  desirable if field-controlled SHG is to be achieved
  \cite{Sirtori92}. For example, \stwo\ is designed to be
  doubly-resonant at $F=43\,\mbox{kV/cm}$; when the field is switched
  off, the intersubband gaps become $0.132\,\mbox{eV}$ ($1\rightarrow
  2$) and $0.103\,\mbox{eV}$ ($2\rightarrow 3$), i.e., the resonance
  condition is lost; accordingly, the SHG drops by two orders of
  magnitude, although the value of $\mu$ is barely affected.

  The benefit from huge non-linearities which can be obtained by
  quantum confinement can be limited by absorption from the material
  at the SHG frequencies; it has been suggested \cite{Rosencher93}
  that optimal {\em conversion} efficiency might be obtained with
  $\mu_{12}/\mu_{13}\simeq 1$. On the other hand, our simulations show
  that optimal {\em quantum} (i.e., confinement-induced) SHG
  efficiency is generally obtained with $\mu_{12}/\mu_{13}\simeq 2$
  (this is, e.g., the case for devices \sone\ and \stwo), which
  reduces the conversion efficiency by a factor of $\sim 2$.

  In summary, the present investigation of the optimal composition for
  double-resonant SHG in GaAs-based devices suggests that the largest
  quantum confinement-induced SHG is obtained in asymmetric
  double-quantum wells with narrow and high barriers and a graded
  composition sideways to the largest well. Furthermore, an electric
  field does not necessarily improve the SHG if structural parameters
  are optimized.

  I acknowledge several useful discussions with Fausto Rossi.

\newpage
\widetext

  \begin{table} 
  \label{tab:best} 
  \caption{Compositional parameters for the optimized devices \sone\
  and \stwo. \sone\ is obtained with $F=0$, while \stwo\ is obtained
  including $F$ in the optimization. For each structure, the width of
  each layer, $n$ (in ML), and the corresponding Al concentration, $x$
  is reported; two capping layers with $x=0.4$ enclose each
  structure. Inter-subband dipole moments, $\mu_{ij}$, and $\mu =
  |\mu_{12}\mu_{23}\mu_{31}|$ are also reported in the
  table. Calculations have been performed with $\hbar\omega=
  0.116\,\mbox{eV}$ and $\Gamma=7\,\mbox{meV}$; the resulting
  inter-subband gaps match $\hbar\omega$ within 1 meV.  The
  correponding potential profiles and subband charge densities are
  shown in Fig.\ 1.}
  \begin{tabular}{rcccccccccccccc} 
  \sone\ & \multispan{13}{$(F = 0)$ \hfill} \\\hline 
  $n$ : & (cap layer) &   6   &   12   &   2   &   16   &   6   &   6   & (cap layer) & & & \\ 
  $x$ : &   0.400     & 0.149 & 0.130 & 0.363 & 0.000 & 0.286 & 0.393 &    0.400    & & & \\\hline 
  \multispan{14}{ $\mu = 3.27\,\mbox{nm$^3$}$  \hspace{4truemm} 
                  $\mu_{12} = 1.61\,\mbox{nm}$ \hspace{4truemm} 
		  $\mu_{23} = 2.41\,\mbox{nm}$ \hspace{4truemm}
		  $\mu_{31} = 0.84\,\mbox{nm}$ \hspace{4truemm} \hfill} \\\hline\hline
  \stwo\ & \multispan{13}{$(F = 43\,\mbox{kV/cm})$ \hfill} \\\hline
  $n$ : & (cap layer) &   4   &   6   &   4   &   14   &   4   &   4   &   4   &    6   & (cap layer) \\ 
  $x$ : &   0.400     & 0.385 & 0.328 & 0.203 & 0.000 & 0.209 & 0.195 & 0.143 &  0.121 &   0.400     \\\hline 
  \multispan{15}{ $\mu = 3.13\,\mbox{nm$^3$}$  \hspace{4truemm} 
                  $\mu_{12} = 1.63\,\mbox{nm}$ \hspace{4truemm}
		  $\mu_{23} = 2.38\,\mbox{nm}$ \hspace{4truemm} 
		  $\mu_{31} = 0.81\,\mbox{nm}$ \hspace{4truemm} \hfill} \\
  \end{tabular}
  \end{table}

  \begin{figure} 
 \noindent
 \unitlength1mm
 \begin{picture}(100,110) 
 \put(30,0){\psfig{figure=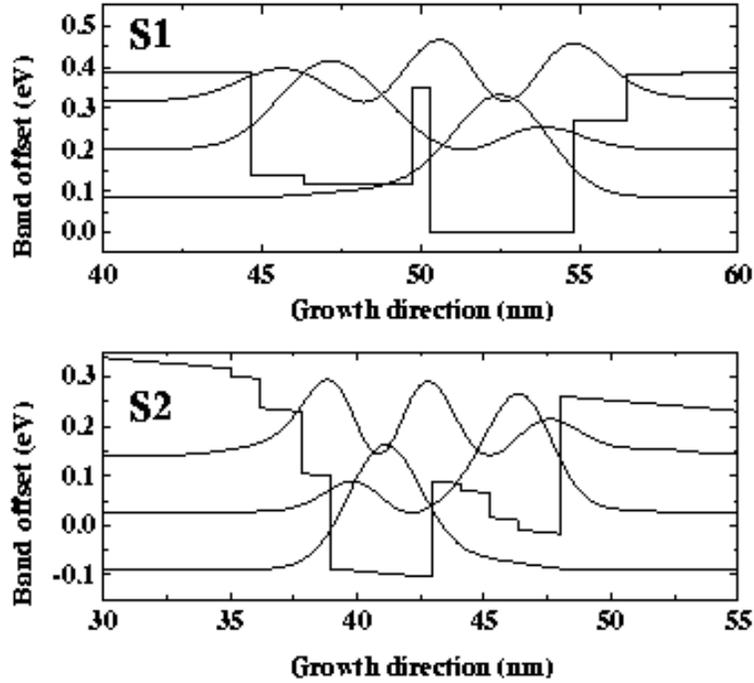,width=110mm}} 
 \vspace{6truemm}
 \end{picture}
  \label{fig:best} 
  \caption{Potential profiles of the
  optimized devices \sone\ and \stwo; the corresponding compositions
  are detailed in Tab.\ 1. Also shown are the charge densities of the
  three bound states involved in the SHG, vertically shifted by the
  confinement energy.} 
  \end{figure}

\newpage

  \begin{figure} 
 \noindent
 \unitlength1mm
 \begin{picture}(100,110) 
 \put(30,0){\psfig{figure=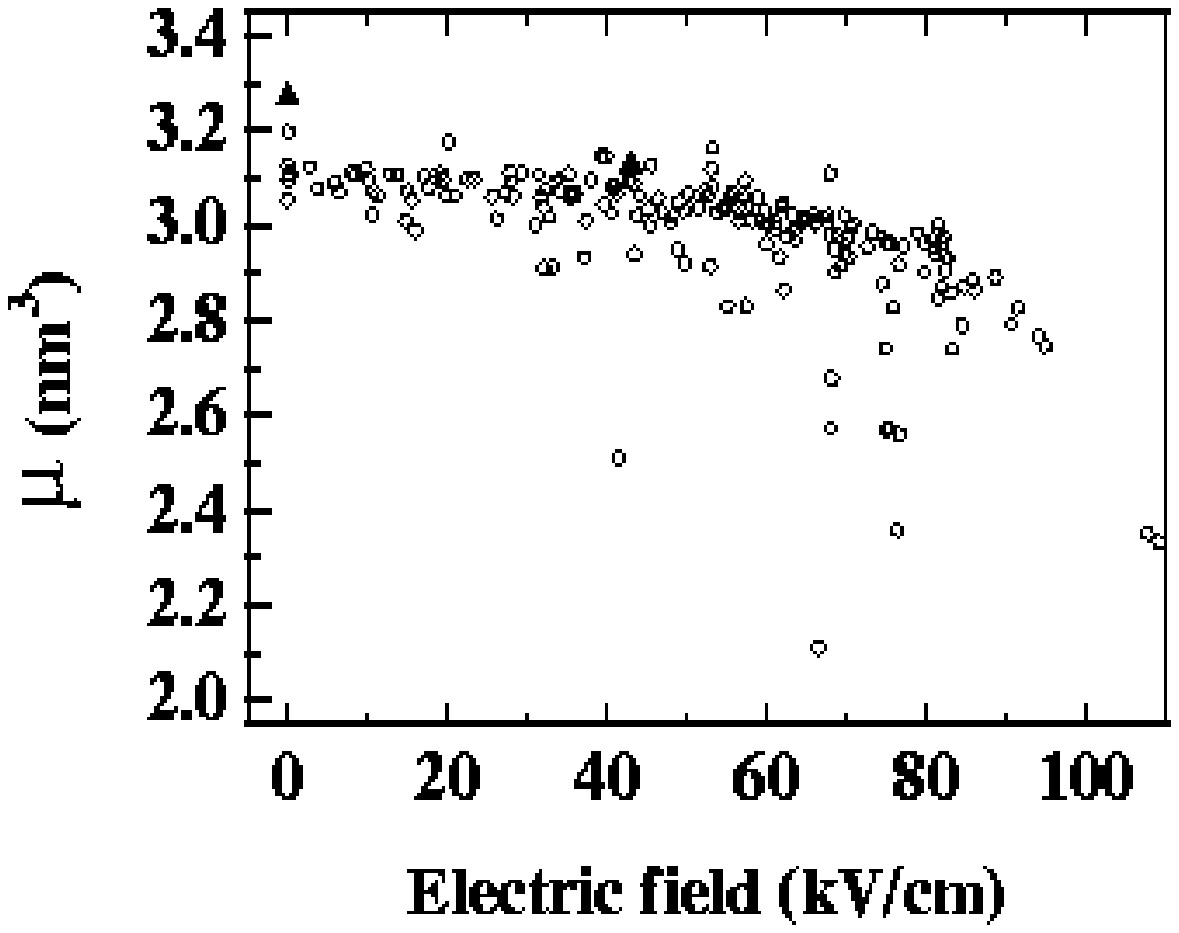,width=110mm}} 
 \vspace{6truemm}
 \end{picture}
  \label{fig:mu-vs-field} 
  \caption{Dipole moment $\mu$
  vs electric field for a set of 200 devices (open dots) which have
  been optimized by evolutionary dynamics with respect to
  compositional parameters and electric field. Also shown are the
  values corresponding to devices \sone\ and \stwo\ (full triangles).} 
  \end{figure}


\begin{references}

  \bibitem[*]{byline1} E-mail: goldoni@unimo.it

  \bibitem{Rosencher89} E. Rosencher, P. Bois, J. Nagle, E. Costard,
  and S. Delaitre, Appl. Phys. Lett. {\bf 55}, 1597 (1989).

  \bibitem{Seto94} M. Seto, M. Helm, Z.Moussa, P. Boucaud,
  F. H. Julien, J.-M. Lourtioz, J. F. N\"utzel, and G. Abstreiter,
  Appl. Phys. Lett. {\bf 65}, 2969 (1994).


  \bibitem{Boucaud90} P. Boucaud, F. H. Julien, D. D. Yang, and
  J-M. Lourtioz, E. Rosencher, P.  Bois, and J. Nagle,
  Appl. Phys. Lett. {\bf 57}, 215 (1990).

  \bibitem{Rosencher91} E. Rosencher and Ph. Bois, Phys. Rev. B {\bf
  44}, 11315 (1991).

  \bibitem{Bewley93} W. W. Bewley, C. L. Felix, J. J. Plombon,
  M. S. Sherwin, M. Sundaram, P. F. Hopkins, and A. C. Gossard,
  Phys. Rev. B {\bf 48}, 2376 (1993).

  \bibitem{Rosencher93} E. Rosencher, J. Appl. Phys. {\bf 73}, 1909
  (1993).

  \bibitem{Sirtori91} C. Sirtori, F.  Capasso, D. L. Sivco, S. N. G. Chu,
  and A. Y. Cho, Appl. Phys. Lett.  {\bf 59}, 2302 (1991).

  \bibitem{Sirtori92} C. Sirtori, F. Capasso, D. L. Sivco, A. L.  Hutchinson, 
  and A. Y. Cho, Appl. Phys. Lett. {\bf 60}, 151 (1992).  

  \bibitem{Liu00} A. Liu, S.-L. Chuang, and C. Z. Ning,
  Appl. Phys. Lett. {\bf 76}, 333 (2000).

  \bibitem{Khurgin} J. Khurgin, Appl. Phys. Lett.{\bf 51}, 2100 (1987);
  Phys. Rev. B {\bf 38}, 4056 (1988).

  \bibitem{Kuwatsuka94} H. Kuwatsuka and H. Ishikawa, Phys. Rev. B
  {\bf 50}, 5323 (1994).

  \bibitem{Park99} T. Park, G. Gumbs, and Y. C. Chen,
  J. Appl. Phys. {\bf 86}, 1467 (1999).

  \bibitem{Shaw93} M. J. Shaw, K. B. Wong, and M. Jaros, Phys. Rev. B
  {\bf 48}, 2001 (1993).

  \bibitem{Fejer89} M. M. Fejer, S. J. B. Yoo, R. L. Byer, A. Harwit,
  and J. S. Harris, Jr., Phys. Rev. Lett. {\bf 62}, 1041 (1989).

  \bibitem{Tsang88} L. Tsang, E. Ahn, and S. L. Chuang,
  Appl. Phys. Lett.  {\bf 52}, 697 (1988).

  \bibitem{Tsang92} L. Tsang and S. L. Chuang, Appl. Phys. Lett. {\bf
  60}, 2543 (1992).

  \bibitem{Tomic97} S. Tomi\'c, V. Milanovi\'c, and Z. Ikoni\'c, Phys.
  Rev. B {\bf 56}, 1033 (1997).

  \bibitem{Tomic98} S. Tomi\'c, V. Milanovi\'c, and Z. Ikoni\'c, J.
  Phys.: Condens. Matter {\bf 10}, 6523 (1998).

  \bibitem{Goldoni00} G. Goldoni and F. Rossi, Optics Lett. {\bf 25},
  1025 (2000).

  \bibitem{EP} Z. Michalewicz, {\em Genetic Algorithms + Data
  Structures = Evolution Programs} (Springer-Verlag, Berlin, 1992);
  D. E. Goldberg, {\em Genetic Algorithms in Search, Optimization, and
  Machine Learning}, Addison-Wesley (1989); M. Gen and R. Cheng, {\em
  Genetic algorithms \& Engineering Design} (John Wiley \& Sons, Inc.,
  New York, 1997).


  \bibitem{modified-qdoes} The alghoritm described in
  Ref. \onlinecite{Goldoni00} has been modified to include the field
  in the evolutionary dynamics.


  \bibitem{BO} The band gaps are chosen as follows: the \AlGaAs\
  band-gap $E_g(x)$ is obtained from $E_g(x) = E_g^{\mbox{\scriptsize
  GaAs}}+1.36x+0.22x^2$ [C. Bosio {\em et al.}, Phys. Rev. B {\bf 38},
  3263 (1988)]; the conduction band offset is obtained subtracting the
  valence band contribution $\Delta E_v(x) = 0.48 x$ [E. T. Yu {\em et
  al.}, Solid State Phys.  {\bf 46}, 2 (1992)].

  \bibitem{EM} The space-dependent effective mass is $m_e(x) =
  0.067+0.083 x$ [Landolt-B\"ornstein, {\em Semiconductors: physics of
  goup IV elements, and III-V compunds}, O. Madelung,
  Ed. (Springer-Verlag, Berlin, 1982), vol. 17].

  \bibitem{NP} We choose $m_e(x,E) = m_e(x) [1+(E-U(x))/E_g(x)]$ for
  $E>U(x)$ and $m_e(x,E) = m_e(x) [1+(U(x)-E)/E_g(x)]$ for $E<U(x)$
  [D. F. Nelson, R. C.  Miller, and D. A. Kleinman, Phys. Rev. B {\bf
  35}, 7770 (1987)].


\end{references}
\end{document}